\documentclass[structabstract]{aa}
\usepackage{graphicx}
\usepackage{lscape}
\usepackage{txfonts}
\usepackage{natbib}
\bibpunct{(}{)}{;}{a}{}{,} 
\usepackage[normalem]{ulem}
\usepackage{rotating}
\usepackage[usenames]{color}

\begin{document}
   \title{A search for new hot subdwarf stars by means of Virtual Observatory tools}


   \author{R. Oreiro
          \inst{1}
\and
	    C. Rodr\'\i guez-L\'opez   
          \inst{2,3}
\and
           E. Solano
          \inst{4}
\and
           A. Ulla
           \inst{3}
\and
          R. \O stensen
	   \inst{5}
\and
            M. Garc\'{\i}a-Torres
           \inst{6}
          }

\institute{Instituto de Astrof\'\i sica de Andaluc\'\i a, Glorieta de la Astronom\'\i a s/n, 18009, Granada,
Spain\\
              \email{roreiro@iaa.es}
	   \and
              University of Delaware. Department of Physics and  Astronomy. 217 Sharp Lab, Newark DE 19716, USA \\
              \email{cristinatrl@uvigo.es}
          \and
             Dpto. de F\'\i sica Aplicada, Universidade de Vigo. Campus Lagoas-Marcosende, 36310, Vigo, Spain \\
         \email{ulla@uvigo.es}
         \and Centro de Astrobiolog\'{\i}a (INTA-CSIC), Departamento de Astrof\'{\i}sica, PO BOX 78, E-28691 Villanueva
de la Ca\~nada, Madrid, Spain. Spanish Virtual Observatory \\
              \email{esm@cab.inta-csic.es}
	  \and
	  Instituut voor Sterrenkunde, K. U. Leuven, Celestijnenlaan 200D, 3001, Leuven, Belgium\\
	      \email{roy@ster.kuleuven.be}
         \and
             \'Area de Lenguajes y Sistemas Inform\'aticos, Un. Pablo Olavide, Crtra de Utrera Km 1, 41013,
Sevilla, Spain\\
             \email{mgarciat@upo.es}
          }


   \date{Accepted}


  \abstract
   {Recent massive sky surveys in different bandwidths are providing new opportunities to modern astronomy. The Virtual 
Observatory (VO) provides the adequate framework to handle the huge amount of information available and filter out data
according to specific requirements.}
   {Hot subdwarf stars are faint, blue objects, and are the main contributors to the far-UV excess observed in elliptical galaxies. 
They offer an excellent laboratory to study close and wide binary systems, and to scrutinize their interiors through
asteroseismology, as some of them undergo stellar oscillations. However, their origins are still uncertain, and
increasing the number of detections is crucial to undertake statistical studies. In this work, we aim at defining 
a strategy to find new, uncatalogued hot subdwarfs.}
   {Making use of VO tools we thoroughly search stellar catalogues to retrieve multi-colour photometry and astrometric 
information of a known sample of blue objects, 
including hot subdwarfs, white dwarfs, cataclysmic variables and main sequence OB stars. We define a procedure to discriminate
among these spectral classes, particularly designed to obtain a hot subdwarf sample with a low contamination factor.
In order to check the validity of the method, this procedure is then applied to two test sky regions: the
\textit{Kepler} FoV and to a test region of 300~deg$^2$ around ($\alpha$:225, $\delta$:5)\,deg.}
   {As a result of the procedure we obtained 38 hot subdwarf candidates, 23 of which had already a spectral
classification. We have acquired spectroscopy for three other targets, and four additional ones have an available
SDSS spectrum, which we used to determine their spectral type. A temperature estimate is provided for the
candidates based on their spectral energy distribution, considering two-atmospheres fit for objects with clear infrared
excess, as signature of the presence of a cool companion. Eventually, out of 30 candidates with spectral classification,
26 objects were
confirmed to be hot
subdwarfs, yielding a contamination factor of only 13\%. The high rate of success demonstrates the validity of the
proposed strategy to find new uncatalogued hot subdwarfs. An application of this method to the entire sky will be
presented in a forthcoming work.}
   {}

   \keywords{stars:early-type --
                subdwarfs --
                astronomical data bases: miscellaneous -- Virtual Observatory --
               }

   \maketitle
%

\section{Introduction}

Hot subdwarf stars (hot sds), objects with temperatures exceeding 19\,000~K and 
$\log g\geq$~5, are considered to be the field counterparts of the extended horizontal-brach (EHB) stars found in 
globular clusters. Such an evolutionary state implies a $\sim0.5~M_{\odot}$ canonical mass,
and a core He-burning structure with a very thin H-envelope (M$_{env}$ $\leq$~0.02$M_{\odot}$,
\citealt{heber86}).
This structure prevents them from ascending the asymptotic giant branch (AGB) and, 
once the core helium is exhausted, they evolve towards hotter temperatures before reaching degeneracy and 
cooling as a normal white dwarf star \citep{dorman93}.

Whereas the role of hot subdwarfs as white dwarf progenitors is well understood, the circumstances 
that lead to the removal of all but a tiny fraction of the hydrogen envelope, at about the same time as the core has achieved 
the mass required for the He flash ($\approx 0.47~M_{\odot}$), are still a matter of debate. Two
main scenarios have been proposed to explain the formation of these objects: i) enhancement of the mass 
loss efficiency near the red giant branch (RGB) tip \citep{dcruz96}; ii) mass transfer through
binary interaction \citep{mengel76}.

A large fraction of hot subdwarfs are observationally found in binary systems \citep{ulla98,
maxted01}, supporting the mass transfer scenario.
However, a non negligible percentage appear as single hot sds. From a single evolution point of view, it is unknown,
though, what could cause an enhanced mass loss at the RGB. Binary population synthesis by
\cite{han02,han03} show that common envelope ejection (CEE), stable Roche
lobe overflow and helium white dwarfs merging can produce hot sds in close or wide
binaries, as well as single sds. 
These different paths produce distinct orbital period distributions and a variety of mass ranges for the 
hot sds and binary companions. A confrontation of theory and observations could be the key to clarify the hot sds 
evolutionary state and, moreover, help to fine-tune the processes related to the CEE. Some attempts have already been
made
in this direction \citep{morales03,stroer07}, however, the observed orbital period distribution is
biased towards the 
close binary systems (with higher radial velocity variations), and the mass distribution is far from being
statistically significant. Besides, the distribution of the companion spectral type also suffers from biases associated 
to the catalogue from which they were selected \citep[see e.g.][]{wade09}. 

The total mass of some hot sds could be inferred for the few eclipsing binary systems known
\citep{ostensen10,for10}. Thanks to the existence of stellar pulsations
in some B-type hot sds (sdBs), the total mass could also be determined using asteroseismic tools for
a handful of cases
\citep[and references therein]{randall09}.
Unfortunately, there are few short-period sdB pulsators with enough excited modes to perform asteroseismic
analysis. 
However, this may be overcome soon, as we are entering the age of space based asteroseismology. Missions like CoRoT
\citep{auvergne09} and {\it Kepler} \citep{borucki10} have eventually opened the door
to asteroseismology of the long-period sdB pulsators \citep{charpi10,vangrootel10}, which are more numerous than
short-period ones, 
but far more challenging for ground-based observations.

The exciting discovery of the first O-type pulsating hot sd \citep[sdO,][]{woudt06} has been somehow tarnished
by the fact that no other similar objects
have been found up to now. In spite of the extensive searches that have been performed \citep{cris07}, new discoveries are hindered by the
lack of catalogued sdOs in the temperature range of the unique pulsator, $T_\mathrm{eff} \sim$~70\,000~K
\citep{fontaine08,crl10}. O-type hot subdwarfs are less numerous in general than sdBs and, in particular, only
about 40 catalogued sdOs have a temperature estimate within a $\pm$5\,000\,K box around the unique sdO pulsator. Much
fewer match a similar $\log g$ and helium abundance.

The scarcity of catalogued sdOs at this temperature range may in part be attributed to the
historical difficulty in obtaining NLTE model atmosphere grids with
$T_{\rm eff}$ values over 60\,000~K, firstly overcome by \citet{dreizler90}. Since then, only a few quantitatively
significant
spectral analysis of sdOs were undertaken \citep{thejll94,bauer95,stroer07}.
The deficit in pulsating sdOs may be due to sdOs having different evolutionary channels and/or large chemical
inhomogeneities. \cite{stroer07} found that sdOs with
subsolar He abundances never showed C or N lines and were scattered in the HR diagram. On the other hand, sdOs with
supersolar He abundances always showed C and/or N lines and similar parameters around $T_\mathrm{eff}
\sim$~50\,000~K and $\log g \sim~5.5$. 

Increasing the number ratio of hot sds in the different galactic populations may help to sort out their 
origins, as suggested by \cite{altmann04}: the binary scenario of \cite{mengel76} would be favoured if the ratio of sdBs
in the halo, thin and thick disk is similar to that of other evolved stars; on the contrary, if the extensive mass loss
scenario
proposed by \cite{dcruz96} is dominant, sdBs in the disk should be more numerous compared to other mid-temperature
HB stars. Some light could be
shed on the formation processes if the low number of known halo and thin disk sdBs risen. Note that, as
described in Section 2, most surveys for faint blue targets intentionally avoided the galactic disk to diminish
contamination with OB stars.

The aim of this work is to devise a procedure to identify new hot sds obtaining the purest possible sample. Our main 
ally will be the Virtual Observatory\footnote{http://www.ivoa.net} (VO) which is becoming an essential, thourough,
time-saving tool in aid of the overwhelmed-by-data astronomers. Modern observational astronomy profit from large area,
multiwavelength surveys, whose data are stored in different archives and formats. Although data can be queried through
different access
methods, the lack of interoperability among astronomical services can hinder to get the most out of combined information from several 
surveys. These drawbacks can can be overcome if we work in the framework of
the Virtual Observatory, an 
international initiative designed to provide the astronomical community with the data access and the research tools
necessary to enable the exploration of the digital, multi-wavelength universe,
resident in the astronomical data archives.

We make use of VO tools throughout the paper to get advantage of an easy data access and analysis for our scientific
purpose. In what follows, 
we overview the conventional catalogues used to select hot subdwarfs in Section~2. 
In Section~3 we describe our devised method to search for hot sds. In Section~4 we present an application of the
method to the {\it Kepler} FoV and a test region as well as the results obtained from the spectroscopic 
follow-up of our list of candidates. In Section 5 we pay attention to binary hot sds candidates. Finally, in Section 6 we summarize our findings.



\section{Hot Subdwarfs Surveys}

Surveys in search for faint blue stars began and flourished around the 60's. \citet{greenstein60} gathered under the
term {\it Faint Blue Stars} all not-well understood spectra of stars that in the HR diagram sat below the main
sequence. Intense
surveys for new blue subluminous stars followed the pioneer discoveries of \citet{humason47}, among them:
\citet{feige58} searched for faint blue stars brighter that $B_{pg}$=14~mag, within 6\,000~deg$^2$ around both galactic
poles
and found 114 objects, later spectroscopically analysed by \cite{sargent68}. \citet{haro62} published the {\it Faint
Blue Stars near the South Galactic Pole}, a catalog with about 8\,700 stars, based on Johnson photometric indexes, up to
magnitude 19, comprising hot sds, white dwarfs (WDs) and quasars, and for which photometric indexes,
spectroscopic and proper motion data were given. \citet{greenstein66} performed a spectroscopic study of about a hundred
faint blue stars mainly at the galactic poles, from the catalogues of \citet{humason47}, \citet{iriarte57},
\citet{chavira58} and \citet{feige58}, which allowed for the first time to distinguish between hot sds, WDs and halo or
horizontal branch stars. The most recent catalogues of hot sds used nowadays are the following:

\subsection{The Palomar-Green (PG) Catalog of Ultraviolet-Excess Stellar Objects}

\cite{green86} $U-B$ photographic survey lists about 1\,900 objects with a limiting magnitude B$_{pg}$=16.7~mag
covering
about 11\,000~deg$^2$ at Galactic latitudes $|b|>30^\circ$ and declinations $\delta < -10^\circ$. A total of 1715
objects showing
ultraviolet excess, given by $(U-B)_{pg}< -0.46$ were observed spectroscopically for classification. This yielded over
900 hot sds, which makes up $\sim$53\% of the catalogue objects.

\subsection{The Kitt Peak-Downes (KPD) Survey for Galactic Plane Ultraviolet-Excess Objects}

\citet{downes86} found 60 hot sds ($\sim$40\% of the objects) and 10 WDs from a 1\,000~deg$^2$ two colour
photographic and spectroscopic survey of
the Galactic plane, obtaining spectra for about 700 UV-excess candidates. Accurate space densities could be determined
for the first time for Galactic plane UV-excess objects (i.e. hot sds, WDs and cataclysmic variables), due to the homogeneity of the
sample, which was complete to B$_{pg}$=15.3~mag.

\subsection{The Montreal-Cambridge-Tololo (MCT) Survey of Southern Subluminous Blue Stars}

The Montreal-Cambridge-Tololo photographic and spectroscopic Survey of southern subluminous blue stars
\citep{lamontagne00,demers86} covers $\sim$6800~deg$^2$ centered on the south Galactic polar cap, at latitudes
below $b$=-30$^\circ$ not covered by the PG survey, and being complete down to B$_{pg}$=16.5~mag. The criteria for
selecting
candidates was $(U-B)_{pg} \leq -0.6$, leading to some 3\,000 objects, for a third of which, spectroscopy was performed.
Results for the analysis of the region of $\sim$800~deg$^2$ of the south Galactic cap are given: of 188
objects, 40\% were found to be hot sds.

\subsection{A Catalogue of Spectroscopically Identified Hot Subdwarfs}

\cite{kilkenny88} made the considerable effort of collecting 1225 known hot sds spectroscopically identified from
different literature sources and,
in that moment, yet to
be published data. The main sources for this compilation are the PG and KPD surveys. 
This was the most extensive hot sds catalogue until the release of {\it The Subdwarf Database} \citep[][see
below]{ostensen06}.

\subsection{The Hamburg-Schmidt/ESO (HQS/HES) Quasar Survey}

{\it The Hamburg-Schmidt Survey} \citep{hagen95}, with the prime scientific goal of providing new, bright QSOs, have
also been the source of new hot sds. The prime scientific goal of the northern survey ($\sim$14\,000~deg$^2$) is to
provide a complete sample of bright, high-redshift QSOs, to
expand the PG-Survey in area and depth up to B$<$17~mag. The quasar search was extended to the southern sky \citep[{\it
The Hamburg/ESO Survey},][]{wisotzki91,wisotzki94,wisotzki00}, where it aims at covering $\sim$5\,000~deg$^2$
for sources with B$<$16.5~mag. In addition, the digitized data base is currently used in the search for hot stars by the
Hamburg-Bamberg-Kiel collaboration \citep[see e.g.][]{heber91,edelmann03}. Candidate hot stars, for which a follow up
and later analyses were performed, were selected on the basis of bluest spectra and visual classification. A initial
candidate list of 400 objects yielded $\sim$50\% hot sds \citep{edelmann03}.

\subsection{Edinburgh-Cape (EC) Survey}
The {\it Edinburgh-Cape Survey} \citep{stobie97,kilkenny97,kilkenny10} aim is to discover blue stellar objects brighter
than B$\sim$18 in southern sky Galactic latitudes $|b|>30^\circ$ and declination $\delta < -12.5$, meaning
$\sim$8\,000~deg$^2$.

The criterium to select blue stellar objects from UK Schmidt telescope plates is $(U-B)_{pg} < -0.4$. The survey was
divided in 6 zones, each comprising $\sim$~1500~deg$^2$. The first release of the survey \citep{kilkenny97} gives
results for the analysis of Zone~1, yielding 675 hot blue objects with a 45\% of hot sds. For the most
up-to-date
status of the project, we refer the reader to \citet{kilkenny10}

\subsection{SPY - The ESO Supernova type Ia Progenitor Survey}
SPY \citep{napi03} is a survey designed to search for short period binary WDs as potential progenitors of type Ia
supernovae. The SPY obtained accurate radial velocities for WD candidates brighter than B$=$16.5~mag belonging to a
variety of source catalogues, mainly the \citet{mccook99}, but also the HQS/HES, MCT, and EC catalogues. \cite{napi04}
found 46 sdBs and 23 sdOs due to misclassifications in the input catalogs.

\subsection{The Subdwarf Database}

{\it The Subdwarf Database}\footnote{http://www.ing.iac.es/ds/sddb/} \citep{ostensen06} is
the latest compilation of any object ever classified
as a hot subdwarf. Initially based on the compendium by~\cite{kilkenny88}, it is in a continuous process of up-dating,
and nowadays contains more than 2400 entries. For each entry, the database provides links to
finding charts, the SIMBAD
Astronomical Database, and data available in the literature, namely $T_\mathrm{eff}$, surface gravity, helium abundance,
photometry and spectral classification. A quality flag is given for the derived spectral classes, to give an estimation
of the reliability of the determination.

Given that {\it The Subdwarf Database} is the most complete compilation of hot sds, it is of unvaluable help in any
study aiming at detect any yet unclassified hot subdwarf, as we are attempting here.

\section{The search method}
\label{sec:procedure}

The main objective of this work is to design a procedure to identify new hot subdwarfs.
Special care is made to avoid contamination from other types of objects as much as possible, giving more
importance to the successful rate (low contamination
factor) than to the completeness of the sample of new hot sds found. Former faint blue star catalogues,
used to select hot sds, 
had also a high percentage of WDs. To name but a few: the PG catalogue had $\sim$~50\% of hot sds, but featured a
25\% of WDs; the MCT survey showed $\sim$~40\% cent of hot sds and $\sim$~30\% of WDs; whereas the EC
survey of Zone~1 showed 45\% of hot sds and 15\% cent of WDs.

Since we aim at obtaining a subdwarf candidate sample as pure as possible, we define the best strategy using
spectroscopically classified bona-fide catalogues: 

\begin{itemize}
 \item {\it The Subdwarf Database} \citep{ostensen06} as the hot sds sample. 
\item {\it SDSS4 confirmed White Dwarf catalogue} \citep{eisenstein06} to obtain a list of white dwarfs.
\item The {\it Catalogue of Cataclysmic Variables} \citep{downes06}.
\item The {\it Photometry and spectroscopy for luminous
stars catalogue} \citep{reed05} to obtain main sequence OB stars.
\end{itemize}

We considered WDs, cataclysmic variables (CVs) and OB stars because they have a photometric
signature similar to that of hot subdwarfs and represent, therefore, the main sources of
pollution in our study. Table~\ref{tab:numbers} indicates the number of targets
in each input catalogue.

The methodology proposed makes use of existing data from different surveys. The data gathering is described in
\S~\ref{s:archives}, which is then used to filter out non hot sds: combined {\sc 2mass} and {\sc
galex} photometry (\S~\ref{s:2massgalex}) will reject red targets and a large fraction of contaminators; 
proper motion information (\S~\ref{s:ppm}) will help to discriminate between kinematic populations, and eventually a
temperature estimate given by the fit to the spectral energy distribution (\S~\ref{s:vosa}) will further
improve the selection of hot sds.


\subsection{Archives data gathering}\label{s:archives}

We made use of TOPCAT\footnote{http://www.star.bris.ac.uk/$\sim$mbt/topcat/}, a VO-tool to work with tabular
data, to access and
download {\sc galex}\footnote{http://galex.stsci.edu/GR4/} and {\sc
2mass}\footnote{http://www.ipac.caltech.edu/2mass/} \citep{morrissey07,skrutskie06} photometric magnitudes
for all the sources in
the input catalogues. Only the best coordinate match within 5 arcsec was considered. 

\begin{figure}
   \centering
   \includegraphics[width=6cm,angle=-90]{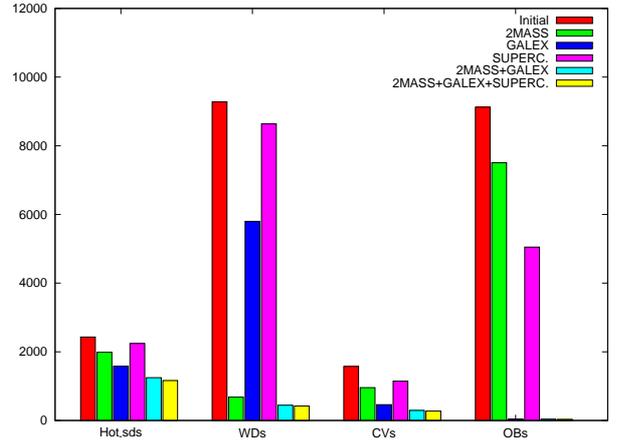}
      \caption{Histogram showing the initial number of objects in the input catalogues, those with a counterpart in the
databases used, and subsequent remaining numbers
as cross-correlation between catalogues are done. Numbers are taken from Table~\ref{tab:numbers}.}
         \label{fig:hist1}
   \end{figure}

We selected only those targets
having both {\sc 2mass} and {\sc galex} photometry, which significantly limited the test sample.
Only a low fraction of the sample WDs has {\sc 2mass} photometry since they are generally too faint at these
wavelengths, while in the case of OB stars, there is {\sc galex}
photometry for very few targets since {\sc galex} does not cover the galactic plane. Table~\ref{tab:numbers} lists
the
number of targets with available photometric magnitudes for every input list (see also Fig.~\ref{fig:hist1}), and
those having photometry in the two surveys (under {\sc 2mass}+{\sc galex}).

\begin{table*}
\caption{Number of initial hot sds, WDs, CVs and OB targets used for defining the procedure in
Section~\ref{sec:procedure}. The number of objects with available data in {\sc galex}, {\sc 2mass} and SuperCOSMOS is
also
included.}
\label{tab:numbers}
\centering
\renewcommand{\footnoterule}{}  
\begin{tabular}{lcccc}
\hline \hline
  &  Hot sds & WDs & CVs & OBs \\
\hline
Initial nums. in cats.  & 2430        & 9277        & 1578        & 9123 \\
{\sc 2mass}                   & 1985 (82\%) & 680  (7\%)  & 956 (61\%)  & 7504 (82\%) \\
{\sc galex}                   & 1578 (65\%) & 5798 (62\%) & 460  (29\%) & 42 (0.46\%) \\
  SuperCOSMOS             & 2243 (92\%) & 8636 (93\%) & 1145 (73\%) & 5049 (55\%)\\
{\sc 2mass}+{\sc galex}             & 1246 (51\%) & 445 (5\%)   & 292  (18\%) & 40 (0.44\%)   \\
{\sc 2mass}+{\sc galex}+SuperC.     & 1162 (48\%) & 421 (4\%)   & 274  (17\%) & 33 (0.36\%)\\
\hline
\end{tabular}
\end{table*}

{\sc galex} and {\sc 2mass} apparent magnitudes (FUV and NUV, and Ks, respectively) were corrected for galactic
extinction using the $E(B-V)$ values from
\cite{schlegel98} and applying the corresponding correction factors by \citet{wyder05} for
FUV and NUV {\sc galex} filters, and \citet{cardelli89} for the {\sc 2mass} Ks filter:


\begin{eqnarray}
FUV_0 = FUV -A_{FUV} = FUV - 8.376 E(B-V)\label{eq:dered} \\ 
NUV_0 = NUV -A_{NUV} = NUV - 8.741 E(B-V)\label{eq:dered2} \\
Ks_0   = Ks - A_{Ks}     = Ks   - 0.114 E(B-V)\label{eq:dered3}
\end{eqnarray}

\noindent where the $0$ subscripts indicate extinction corrected magnitudes.

Moreover, we also downloaded SuperCOSMOS\footnote{http://surveys.roe.ac.uk/ssa/}, \citep{supercosmos}
proper motions for the input list of targets, which will be used to separate different kinematic populations. A high
fraction of the test objects
have catalogued proper motions, as seen in Table~\ref{tab:numbers}. From now on, we will only use targets having
{\sc 2mass} and {\sc galex} photometry and proper motions given by SuperCOSMOS (under {\sc 2mass}+{\sc galex}+SuperC in
Table~\ref{tab:numbers}). This leaves us with a low fraction of WDs and a tiny fraction of OBs with respect to the
original input catalogues, but we also lose about half of the initial hot sds.

%

\subsection{Ultraviolet-infrared colour filter}\label{s:2massgalex}

\citet{rhee06} already presented a selection strategy to identify hot subdwarfs using a combination
of photometric indices
from {\sc galex} and
{\sc 2mass}.
The {\sc galex} satellite is performing a series of sky surveys in two ultraviolet
bands, FUV (1344 -- 1786\,\AA) and NUV (1771 -- 2831\,\AA), whereas {\sc 2mass} scanned the entire sky in
three near-infrared bands, J (1.25 $\mu$m), H (1.65 $\mu$m), and Ks (2.17 $\mu$m).  Given their large area coverage,
the combination of both datasets represents an excellent approach to 
separate blue from red targets. 

\citet{rhee06} cross-matched the {\sc galex} and {\sc 2mass} catalogues in a 3500 deg$^{2}$ region. They proposed as hot
sds
candidates those falling within the limits $-4 < (FUV-Ks) < 1$ and
$-1.5 < (FUV-NUV) < 0.5$ in a two-colour diagram. However, follow-up spectroscopic observations for a subsample
of 34 subdwarf candidates resulted in 60\% contamination from other blue objects.
We will therefore attempt to use slightly refined search criteria.

\begin{figure}
   \centering
   \includegraphics[]{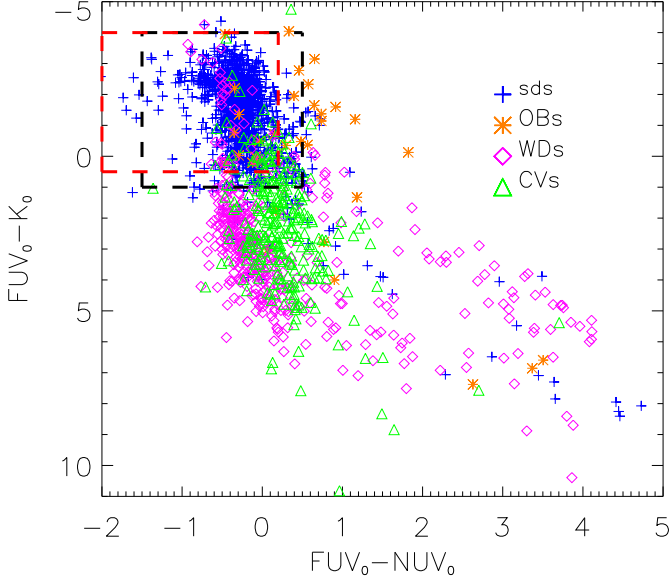}
      \caption{Test sample of hot sds, WDs, CVs and OBs having {\sc 2mass}+{\sc galex}+SuperCOSMOS data. Note
that those hot sds with $FUV_0-NUV-0>2$ all have a $FUV$ magnitude above 18.7, close or above the limiting magnitude,
which is causing their red appearance. The black dashed box
comprises objects within the {\sc galex}-{\sc 2mass} selection limits given by \citet{rhee06}. The red dashed box gives
selection limits proposed in this work. See text and Table~\ref{tab:rhee_crit} for further details.
              }
         \label{fig:2massgalex}
   \end{figure}

Following \citet{rhee06}, we plotted $(FUV_0-Ks_0)$ vs.
$(FUV_0-NUV_0)$ for the input sample (Fig.~\ref{fig:2massgalex}).
The four different object classes under study are plotted with different colours and symbols.
The black dashed box indicates the limits for the hot sds selection proposed by \citet{rhee06}.
It results in a very effective procedure to differentiate hot sds from the other samples:
92\% of the considered hot sds lie within this box, while only 12\% of the
WDs, 24\% of the CVs and 45\% of the OBs fulfill this selection criterion.

Slight changes on the box limits result in a purer or more contaminated hot sds sample, always
at the expense of sacrifying good candidates. After some tests, we redefined Rhee's selection box as:

\begin{eqnarray}
-4 < (FUV_0 - Ks_0) < 0.5\label{eq:fuvk} \\ 
-2 < (FUV_0 - NUV_0) < 0.2\label{eq:fuvnuv}
\end{eqnarray}

\noindent which is a good compromise between obtaining a low contamination factor and
avoiding the rejection of too many hot sds. The new defined selection limits are overploted as a red
dashed box in Fig.~\ref{fig:2massgalex}.
Table~\ref{tab:rhee_crit} compares the percentage of targets surviving
the two different selection criteria. As indicated, the new limits diminish contamination, in particular from
CVs and OBs.

Given that white dwarfs can be nearby objects, their magnitudes might have been overcorrected for galactic
extinction. As a cross-check, if half of the $E(B-V)$ values from
\cite{schlegel98} were used, the resulting $FUV_0-K_0$ colours would be shifted towards blue $\sim0.15$\,mag on
average, and only one new white dwarf would enter the selection criteria indicated in Fig.~\ref{fig:2massgalex}.

%

%

\begin{table}
 \caption{1 \& 2: Percentage of input objects fulfilling the selection criteria of two-colour indices $(FUV_0-Ks_0)$ -
$(FUV_0-NUV_0)$ given by~\citet{rhee06} and in this paper. 3: $H(NUV_0)$ criterion proposed in this paper. Percentages
refer to objects having data in
{\sc 2mass}+{\sc galex}+SuperCOSMOS catalogues (see Table~\ref{tab:numbers} for further details). 4: Absolute number of
objects
fulfilling criteria 2 \& 3. 5: Percentage of objects with $T_{\rm eff}>$19\,000\,K; in parenthesis: considering also
objects with a bad SED fit (see \S\ref{s:vosa}). 6: Same as 5, but in absolute numbers.}
\label{tab:rhee_crit}
\centering
\begin{tabular}{l@{}ccccc}
\hline\hline
& Hot sds & WDs & CVs & OBs \\
\hline
1. Rhee et al. criteria & & & & \\
$-4<(FUV_0-Ks_0)<1$  & 92\% & 12\% & 24\% & 45\% \\
 $-1.5<(FUV_0-NUV_0)<0.5$ &  &  &  &  \\
\hline
2. This paper & & & & \\
$-4<(FUV_0-Ks_0)<0.5$  & 87\% & 10\%  & 13\% & 33\% \\
 $-2  <(FUV_0-NUV_0)<0.2$ &  &  &  &  \\
 \hline                                   
3. This paper & & & & \\
$19< H(NUV_0) < 27$  & 83\% & 3\%  & 11\% & 21\% \\
\hline
4. Total number after 2\&3   & 960 & 14  & 30 & 7 \\
\hline
5. VOSA $T_{\rm eff}>$19\,000 & 64(72)\% & 3(3)\%& 1(4)\%& 3(6)\%\\
\hline
6. Total number after 2,3,\&5   & 749 (846) & 14 (14) & 3 (12) & 1 (2) \\
\hline                                   
\end{tabular}
\end{table}

\subsection{Proper motion filter}\label{s:ppm}
Further improvement to this colour-colour diagram is obtained when reduced
proper motions (RPM) are used to separate populations of stars with different kinematics. The RPM in any filter is
calculated as:
\begin{equation}
H(Filter) = m + 5\log_{10}(\mu)+5
\end{equation}

\noindent where $m$ is the apparent magnitude for the corresponding filter and
$\mu$ the proper motion measured in milliarcsec per year. \citet{salim02} used a combination of optical and
infrared RPMs
to achieve a rather good isolation for the bluest WDs from main sequence stars and (cool) subdwarfs, although
contamination factors were not given. 

\begin{figure}
   \centering
   \includegraphics[]{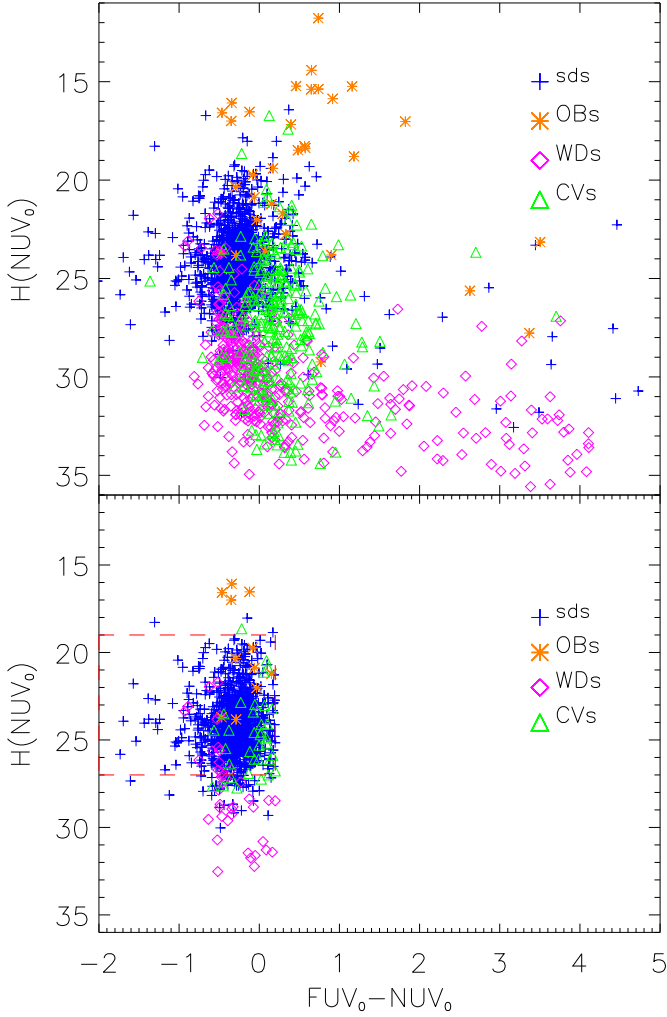}
      \caption{Above: $H(NUV_0)$ against $(FUV_0-NUV_0)$ for the sample in Fig.~\ref{fig:2massgalex}. Notice inverted
y-axis scale. Below: Only objects fulfilling Eqs.~\ref{eq:fuvk} and~\ref{eq:fuvnuv} are plotted (see
\S\ref{s:2massgalex}). The red dashed box gives
$H(NUV_0)$ selection limits proposed in this work. See text and Table~\ref{tab:rhee_crit} for further details.}
         \label{fig:rpm}
   \end{figure}

With this aim, we constructed an UV-RPM diagram for the test catalogues. In the upper panel of Fig.~\ref{fig:rpm} 
we plotted H($NUV_0$) against ($FUV_0-NUV_0$)
for the complete sample. Subdwarfs can be distinguished from OB main-sequence stars as they are 
several magnitudes dimmer at the same colour and typically have larger velocities. These effects 
tend to move hot sds significantly below the OB stars in the reduced proper-motion diagram. White dwarfs, 
with even fainter magnitudes, also appear clearly separated. Cataclysmic variables, on the other hand, quite
overlap with hot
 sds and WDs.

In the lower panel of Fig.~\ref{fig:rpm} only targets fulfilling Eqs.~\ref{eq:fuvk}-\ref{eq:fuvnuv} (see
\S\ref{s:2massgalex}) are included. We now select targets within
$19 < H(NUV_0)<27$, to end up with a sample almost devoid of WDs,
containing only a 3\% of the initial {\sc 2mass}+{\sc galex}+SuperCOSMOS selection, and a
low fraction of the initial CVs (11\%) and OBs (21\%), as
listed in Table~\ref{tab:rhee_crit} (point 3). Absolute numbers after application of these selection criteria are also
included in Table~\ref{tab:rhee_crit} (point 4).

\subsection{Spectral energy distribution fit}\label{s:vosa}

Finally, we obtained a temperature estimate for the surviving targets fitting their spectral energy distribution (SED).
For this purpose, we employed the VO-tool VOSA\footnote{http://svo.cab.inta-csic.es/theory/vosa/} (VO Sed Analyzer),
which allows the user to query photometry from different catalogues, and compute $T_{\rm eff}$ from comparison of the
SEDs with
those derived from a grid of theoretical spectra. 

We used the {\sc 2mass} and {\sc galex} photometry of the targets for the SED fit, as well as any photometric
data existent in other public archives. With only {\sc 2mass} and {\sc galex} photometry it is difficult to get
reliable values of the effective temperature due to: (i) {\sc 2mass} photometry is not sensitive
to $T_\mathrm{eff}$ changes within the typical temperature ranges for hot sds, i.e., large
$T_\mathrm{eff}$ variations may still yield good fits and (ii) {\sc galex} magnitudes are highly dependent on reddening
and moderate errors in $E(B-V)$ 
translate into substantial errors in $T_\mathrm{eff}$. 

\begin{figure}
   \centering
   \includegraphics[]{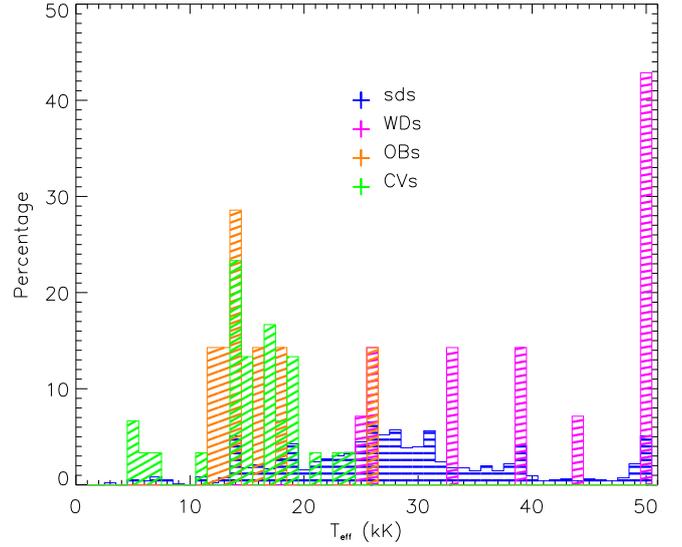}
      \caption{Histogram showing the effective temperature distribution obtained with VOSA for the 960 hot sds (blue),
14
WDs (magenta), 30 CVs (green) and 7 OBs (orange). See text for details.
        }
         \label{fig:hist2}
   \end{figure}

In order to solve this problem we used VOSA to obtain photometry from the SDSS/DR7 \citep{abazajian09}, {\sc
ukidss}
\citep{lawrence07}, CMC-14\footnote{http://www.ast.cam.ac.uk/cmt/cmc14.html} and TYCHO-2 \citep{hog00} services
for the objects fulfilling the two-colour and RPM criteria (see point 4 in
Table~\ref{tab:rhee_crit}). The magnitudes are then transformed to fluxes and deredened using the extinction law by
\cite{fitzpatrick99}. The observed SED is fit with a grid of Kurucz model atmospheres~\citep{castelli97} with
ranges 3\,500$<T_{\rm eff}<$50\,000\,K, 2.5$<\log g<$ 5.5. The temperature of the best fit for every object is
represented in the histogram in Fig.~\ref{fig:hist2}. Note that this figure is shown in percentage
for visibility reasons, given the low number of surviving objects other than hot sds.

We can see in Fig.~\ref{fig:hist2} that WDs have high scattered temperatures, as expected from them being in different
stages of the cooling phase. We note that five WDs have a M dwarf companion, and for all of them we retrieve a $T_{\rm eff}$ estimate 
below 40\,000\,K.  
Concerning OBs and CVs, most of them are within 10--20\,kK. When targets with the
poorest goodness of fit are excluded of the plot, OBs and CVs' location within 10--20\,kK
is emphasized. Therefore, we can use the temperature estimate given by VOSA to discard a large fraction of OB and CV
objects.


We have adopted to discard as hot sds candidates, those objects with $T_{\rm eff}<$19\,000~K and 
an acceptable goodness of fit, while we mantain as
candidates those with $T_{\rm eff}>$19\,000\,K and those with bad fits, no matter their $T_{\rm eff}$. The reason for
this last choice is that O-type hot sds can have effective temperatures exceeding Kurucz's
upper limit of 50\,000~K and the fitting is expected to be poor for these hot objects\footnote{We have checked, however, that for objects with
$T_{\rm eff}$ above the 50\,000\,K Kurucz limit, despite a bad fit is obtained, the VOSA $T_{\rm eff}$ estimate is always above 19\,000\,K.
This ensures that any hot sdO will not be rejected by the selection procedure.}. The implementation in VOSA of a set of NLTE sdO models hotter than
50.000 K is under investigation at the moment, to eventually be used in a further extension of our study.
A poor fit may be also obtained for composite hot sds.

The results are included in Table~\ref{tab:rhee_crit}: 72\% of the initial hot sds survive all the selection
filters
imposed, while only 3\%, 4\% and 6\% of the WDs, CVs and OBs are selected as candidates, which correspond to a very
little fraction of contamination. It is difficult to asess the exact success rate, though, since the initial sample
lists contain quite different number of objects.


\section{Application of the method}\label{s:application}

In the previous section we designed a procedure to select a hot sds sample as pure as possible. In order
to check the
validity of this method, it was applied to two test sky regions
and the results are described below.

\subsection{Test region A: {\it Kepler} FoV}\label{s:testA}

We tested our method in a region of about 420\,deg$^2$, RA:(275,305)\,deg, DEC:(+33,+55)\,deg covering generously
the {\it Kepler} FoV\footnote{http://kepler.nasa.gov/Science/targetFieldOfView/} of 105~deg$^2$.

The {\it Kepler} FoV was chosen to test our method since extensive efforts are being made to select and classify
suitable targets
for long-term photometric monitoring. Within this framework, low-, intermediate- and high-resolution spectra of a large
number of targets in the FoV are being acquired by different groups~\citep{katrien10}. Indeed, we use the {\it
Kepler} test region to be able to confirm our hot sds candidate list, 
thanks to the above mentioned works, without the necessity of performing our own spectroscopic observations.

\begin{figure}
   \centering
   \includegraphics[scale=0.5]{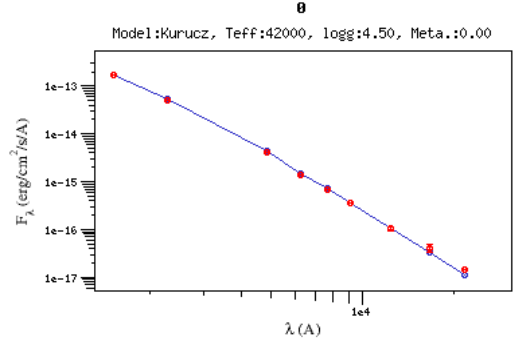}
   \caption{Example of a SED fit performedy by VOSA for object (18:43:07,+42:59:18) in the {\it Kepler} FoV (see
\S~\ref{s:testA} and Table~\ref{tab:candsKepler}).}
         \label{fig:vosa}
   \end{figure}

Our workflow comprised the following steps:

\begin{enumerate}

\item Cross-match:
For each {\sc galex} source within the field, we looked for all {\sc 2mass} counterparts in a 4~arcsec circular
region. If more than one {\sc 2mass}
 counterpart is found, then the {\sc galex} source is removed\footnote{If more than one {\sc 2mass} match are found, the
source is
not further considered. Although the nearest one is supposed to be the right infrared counterpart, we prefer to safely
reject the candidate to avoid any missmatch and any UV contamination from the second close object, given the $\sim$
4.5--6.0~arcsec {\sc galex} point-spread function.}.

\item Data filtering

\begin{itemize}
\item We selected {\sc galex} sources with FUV and NUV values brighter than the 5-$\sigma$ limiting magnitudes (19.9 and
20.8,
respectively). 
\item We dereddened {\sc galex} (FUV, NUV) and {\sc 2mass} (Ks) photometry using Eqs.~\ref{eq:dered}-\ref{eq:dered3} to
obtain the
corrected $FUV_0,
NUV_0$ and $Ks_0$.
\item We selected sources fulfilling $-4 < (FUV_0-Ks_0) < 0.5$ and $-2 < (FUV_0-NUV_0) < 0.2$. This step left us with 90
candidates.
\item We checked if the candidates were already in the catalogues used for defining the procedure
(Section~\ref{sec:procedure}). After
this step, 87 candidates remained\footnote{Three objects are already catalogued: HS1844+5048 in The Subdwarf Database, 
V476\,Cyg in the Catalogue of Cataclysmic Variables and ALS10696 in the OB stars catalogue~\citep{reed05}.}
\item We retrieved SuperCOSMOS proper motions for the candidates using a 5\,arcsec search radius
and applied the RPM selection criteria
($19<H(NUV_0)<27$). This yielded 73 candidates (see Table~\ref{tab:candsKepler73}).
\end{itemize}

\item SED fitting: 
For each candidate, we used VOSA to obtain its effective temperature from the theoretical model that
best fitted the observed SED. In this particular field it was necessary to download the photometry available in the
{\it Kepler} Input Catalog\footnote{http://archive.stsci.edu/kepler/kic10/search.php}(KIC) in order to perform an
acceptable SED reconstruction, since only {\sc galex} and {\sc 2mass} data is retrieved from VO services.  

We thus restricted the analysis to those candidates with $g'r'i'z'$ SDSS photometry obtained during the {\it Kepler} 
preparatory programmes. This requisite limited our list to 21 objects, although we include in
Table~\ref{tab:candsKepler73} the initial 73 candidates for general interest, and indicate the KIC number for these 21
targets with additional photometry in the KIC.

From VOSA analysis, we selected those good fits having $T_{\rm eff}>$19\,000\,K, and those which 
were not correctly fitted regardless their temperature. After this step, we eventually 
ended up with 15 candidates. Table~\ref{tab:candsKepler} gathers their photometry, while Fig.~\ref{fig:vosa} includes
the best SED fit for an example object.

\end{enumerate}

Spectra were already
available for all the candidates in
Table~\ref{tab:candsKepler}~\citep{ostensen10survey,ostensen11}, from which we benefit to check
the validity of our procedure. Thirteen of our 
candidates were classified as hot subdwarfs, the remaining objects being a main sequence B star and a DA star, 
which confirmed the robustness and efficiency of our method (87\%).

Table~\ref{tab:candsKepler} includes the spectral classification for the candidates, besides their corresponding KIC
number. These 15 targets have all been already observed by {\it Kepler}, and Table~\ref{tab:candsKepler} also indicates
the
survey cycle in which they were observed, and a brief description of their light curves.

\subsection{Test region B: RA: 210-240; DEC:0-10}

We also applied our procedure to a region of 300 deg$^2$: RA: (210,240)\,deg, DEC:(0,10)\,deg following the
steps enumerated in the previous section. We
obtained a final list of 11 hot sds candidates, which are included in Table~\ref{tab:candsB2mass}, together with their
{\sc 2mass, galex} and {\sc sdss} photometry, 
along with $T_{\rm eff}$ estimated by VOSA and the spectral classification described below. For these objects we made
use of
Aladin\footnote{http://aladin.u-strasbg.fr/} to
get the spectroscopic and 
catalogue data available in all Virtual Observatory services. The information is shown in Table~\ref{tab:candsB2mass}
under column ``Comments''. 

Given that this region is covered by the Large Area Survey of {\sc ukidss}\footnote{http://www.ukidss.org/},
we repeated the workflow in the same field using the {\sc ukidss} Data Release 7 instead of {\sc 2mass}. 
Its large area coverage (7500 deg$^2$) and depth 
(three magnitudes dimmer than {\sc 2mass}) makes it very adequate to search for non-catalogued, faint hot sds, which
may not be catalogued by {\sc 2mass}. {\sc ukidss}
search yielded 12 additional candidates, 
which are listed in Table~\ref{tab:candsBUKIDSS} with the same information as in Table~\ref{tab:candsB2mass}. 



\begin{table*}
\caption{List of candidates in the {\it Kepler} field of view. FUV, NUV have been taken from the {\sc galex} archive;
$g',r',i',z'$
from
the {\it Kepler} Input Catalogue and J,H,K from {\sc 2mass}. ``U'' means upper limit in the {\sc 2mass} photometry.
$T_{\rm eff}$ is
obtained from the best SED fit performed by VOSA. A bad SED fit is indicated with an (*) under ``Comments''. KIC is the
number from the {\it Kepler} input catalogue. Under ``Comments'' we include: the
{\it Kepler} survey cycle in which each target was observed, a brief description of its light curve: (0): no particular
features; (1): reflecting
binary; (2): $g$-mode pulsator, (3): irregular variable, (4): variability from the companion; and the spectral
classification: (A)
from~\cite{ostensen10survey} or (B): from~\cite{ostensen11} }
\label{tab:candsKepler}
\centering
\renewcommand{\footnoterule}{}  
\begin{tabular}{c@{ }cc@{ }cc@{ }@{ }c@{ }@{ }c@{ }@{ }cc@{ }@{ }c@{ }@{ }c@{ }c@{ }@{ }c@{ }@{ }r@{ }@{ }r@{ }}
\hline \hline
RA      & DEC    & FUV     & NUV     & $g'$ & $r'$ & $i'$ & $z'$ & J & H & K &  $T_{\rm eff}$ &Class. & KIC  &
Comments\\
(J2000) & (J2000)&         &         &   &   &   &   &   &   &   & (VOSA)         &       & number & \\
\hline
18:42:42 & +44:04:06& 16.317 & 16.442 & 17.056 & 17.378 & 17.555 & 17.630 & 16.75 & 16.34 & 15.78 & 25\,000 & sdB  & 8142623 & (*)$Q_1$ (1)(A)\\
18:43:07 & +42:59:18& 14.149 & 14.737 & 15.410 & 15.864 & 16.208 & 16.528 & 16.27 & 16.13 & U:16.24   & 42\,000 &sdO+dM & 7335517 & $Q_3$ (1)(B)\\
18:47:14 & +47:41:47& 13.305 & 13.772 & 14.489 & 14.988 & 15.366 & 15.725 & 15.39 & 15.62 & 15.47 & 41\,000 & He-sdO  & 10449976 & $Q_3$ (0)(B)\\
18:50:17 & +43:58:29& 16.501 & 16.387 & 16.353 & 16.649 & 16.996 & 17.115 & 16.71 & U:17.15 & U:17.00  & 23\,000 & B  & 8077281 & $Q_2$ (3)(A)\\
19:04:35 & +48:10:22& 15.726 & 16.057 & 16.696 & 17.031 & 17.236 & 17.456 & 16.63 & U:16.34 & U:17.11 & 25\,000 &sdB & 10784623 &$Q_5$(1)(B) \\ 
19:05:06 & +43:18:31& 14.208 & 14.391 & 15.058 & 15.516 & 15.894 & 16.193 & 15.81 & 16.06 & U:16.23    & 30\,000 & sdB  &7668647 & $Q_3$ (2)(B)\\
19:08:25& +45:08:32& 15.267 & 15.518 & 16.283 & 16.605 & 16.755 & 16.819 & 16.17 & 15.78 & 15.36 & 26\,000 & sdB &8874184 & (*)$Q_4$(4)(B)\\
19:08:46& +42:38:31& 13.983 & 14.490 & 15.146 & 15.595 & 15.964 & 16.295 & 15.90 & 16.05 & 16.02 & 35\,000 & sdB   &7104168 & $Q_3$ (0)(B)\\
19:09:33& +46:59:04& 14.695 & 15.009 & 15.483 & 15.967 & 16.308 & 16.607 & 16.36 & U:15.71 & U:16.61 &35\,000 & sdB  & 10001893 & $Q_3$ (2)(B)\\
19:10:00& +46:40:24& 14.119 & 14.496 & 14.572 & 14.583 & 14.613 & 14.672 & 13.93 & 13.80 & 13.73 & 18\,000 & sdO+F/G &9822180 & (*)$Q_2$(4)(A)\\
19:14:28& +45:39:09& 15.002 & 15.339 & 15.895 & 16.174 & 16.297 & 16.423 & 16.02 & 15.57 & 15.68 & 26\,000 & sdB  & 9211123 & $Q_3$ (0)(B)\\
19:16:12& +47:49:16& 15.291 & 15.413 & 15.269 & 15.278 & 15.339 & 15.351 & 14.62 & 14.42 & 14.27 & 15\,000 & sdB+F/G &10593239 & (*)$Q_{2,5}$(4$?$)(A)\\
19:26:51& +49:08:48& 14.981 & 14.813 & 15.396 & 15.522 & 15.565 & 15.595 & 14.83 & 14.52 & 14.37 &21\,000 & sdB+F/G & 11350152 & (*)$Q_3$ (4)(B)\\
19:40:32& +48:27:23& 13.544 & 13.969 & 14.093 & 14.175 & 14.269 & 14.369 & 13.68 & 13.56 & 13.56 & 20\,000 & sdB+F/G & 10982905 & (*)$Q_2$ (4)(A)\\
19:43:44& +50:04:38& 13.228 & 13.888 & 14.455 & 14.938 & 15.320 & 15.682 & 15.37 & 15.36 & 15.18 & 50\,000 & DA   &11822535 & (*)$Q_2$ (0)(A) \\
\hline
\end{tabular}
\end{table*}

\begin{table*}
\caption{List of candidates in the Test Region B obtained from the {\sc galex}-{\sc 2mass}-SuperC cross-match. FUV, NUV
have been
taken
from the {\sc galex} archive; $u,g,r,i,z$ from the SDSS Data Release 7 and J,H,K from the {\sc 2mass} Point Source
Catalogue.``U''
means upper limit in the {\sc 2mass} photometry. $T_{\rm eff}$ is
obtained from the best SED fit performed by VOSA. Under ``Comments``: known names
are given when possible; (*) indicates a bad VOSA fitting; (1): catalogued in \cite{ostensennot}; (2): classified as DA
by \cite{koester09}.}
\label{tab:candsB2mass}
\centering
\renewcommand{\footnoterule}{}  
\begin{tabular}{c@{ }cc@{ }cc@{ }@{ }c@{ }@{ }c@{ }@{ }c@{ }@{ }cc@{ }@{ }c@{ }@{ }c@{ }c@{ }@{ }c@{ }c}
\hline \hline
RA      & DEC    & FUV     & NUV     & $u$ & $g$ & $r$ & $i$ & $z$ & J & H & K &  $T_{\rm eff}$ &Class. & Comments\\
(J2000) & (J2000)&         &         &   &   &   &   &   &   &   &   & (VOSA)         &  & \\
\hline
14:14:35& +00:12:36 & 14.378 & 14.770 & 15.389 & 15.725 & 16.240 & 16.585 & 16.921 & 16.46 &  U:16.37 & U:16.81  &48\.000&  & FBS 1412+004\\
14:23:40& +00:10:21 & 16.074 & 16.669 & 17.469 & 17.854 & 18.065 & 17.914 & 17.683 & 16.19 & 15.57 & 15.58 & 44\,000& & \\
14:58:06& +08:51:30 & 13.886 & 14.282 & 14.380 & 14.349 & 14.839 & 15.160 & 15.474 & 15.14 & 15.11 & 15.01 & 23\,000 &sdB &  \\
15:10:42& +04:09:55 & 15.838 & 15.962 & 16.541 & 16.810 & 17.249 & 17.498 & 17.692 & 16.72 & U:16.26 & U:16.45 & 31\,000& sdOB &(1) J15104+0409\\
15:16:46& +09:26:32 & 15.857 & 16.285 & 16.854 & 17.019 & 17.295 & 17.452 & 17.641 & 16.57 & 16.26 & U:16.06  & 19\,000 &sdB+F/G&  \\
15:35:10& +03:11:14 & 14.020 & 14.727 & 24.109 & 15.636 & 16.148 & 16.517 & 16.851 & 16.48 & 16.42 & U:17.16 &50\,000 &DA& (*,2)\,WD1532+033\\
15:43:39& +00:12:02 & 16.204 & 16.437 & 16.708 & 16.726 & 17.027 & 17.169 & 17.325 & 16.71 & 16.26 & U:15.44  & 23\,000&sdB& (1) EGGR 491\\
15:45:46& +01:32:29 & 16.297 & 16.690 & 16.901 & 16.673 & 16.580 & 16.525 & 16.512 & 15.62 & 15.10 & 14.99 & 21\,000 &sdB+F/G & (*) \\
15:51:20& +06:49:04 & 15.362 & 15.591 & 15.987 & 15.891 & 15.936 & 15.949 & 15.988 & 15.26 & 14.98 & 14.90 & 8\,000 &sdOB+X & (*,1)\,J15513+0649\\
15:53:33& +03:44:34 & 16.533 & 16.785 & 16.618 & 16.688 & 16.828 & 16.913 & 17.018 & 16.29 & 16.03 & 15.54 & 24\,000 &He-sdOB &\\
15:56:28& +01:13:35 & 15.338 & 15.633 & 15.867 & 15.985 & 16.387 & 16.707 & 16.935 & 16.46 & 16.63 & U:17.02 & 29\,000&sdB & (1) J15564+01131\\
\hline
\end{tabular}
\end{table*}

\begin{table*}
\caption{List of candidates in Test Region B obtained from the {\sc galex}-{\sc UKIDSS}-SuperC cross-match. FUV, NUV have
been
taken
from the {\sc galex} archive; $u,g,r,i,z$ from the SDSS Data Release 7 and Y,J,H,K from the {\sc ukidss} Large Area
Survey
(DR7).
$T_{\rm eff}$ is obtained from the best SED fit performed by VOSA. 
Under ``Comments'': (*) indicates a bad VOSA fitting; (1) catalogued in \cite{mccook99}; (2) classified as
sdB in
\cite{eisenstein06}; (3) classified as sdO in \cite{eisenstein06}}
\label{tab:candsBUKIDSS}
\centering
\renewcommand{\footnoterule}{}  
\begin{tabular}{c@{ }cc@{ }cc@{ }c@{ }c@{ }c@{ }cc@{ }c@{ }c@{ }c@{ }c@{ }c@{ }l}
\hline \hline
RA      & DEC    & FUV     & NUV     & $u$ & $g$ & $r$ & $i$ & $z$ & Y & J & H & K &  $T_{\rm eff}$ &Class&Comments\\
(J2000) & (J2000)&         &         &   &   &   &   &   &   &   &   &   & (VOSA)         & &\\
\hline
14:15:17& +09:49:26& 17.485 & 17.968 & 18.765 & 19.150 & 19.493 & 19.317 & 18.958 & 18.267 & 17.804 & 17.298 & 17.003 &7\,200 &  & (*)\\
14:24:37& +02:34:19& 15.712 & 16.118 & 16.465 & 16.464 & 16.841 & 17.122 & 16.986 & 17.060 & 17.129 & 17.220 & 17.245 &23\,000 & DA & (1) PG1422+028\\
14:34:40& +06:07:03& 15.988 & 16.524 & 17.288 & 17.714 & 18.175 & 18.258 & 18.140 & 17.508 & 17.142 & 16.633 & 16.458 &39\,000 &  & \\
14:39:18& +01:02:51& 16.326 & 16.285 & 16.391 & 16.367 & 16.769 & 17.060 & 17.340 & 16.968 & 17.054 & 17.139 & 17.216 &21\,000 &  sdB & (2) J1439+0102\\
14:47:30& +03:15:06& 16.049 & 16.222 & 16.311 & 16.137 & 16.550 & 16.855 & 17.108 & 16.748 & 16.740 & 16.698 & 16.809 &20\,000 &  & \\
15:02:30& +09:13:57& 16.190 & 16.354 & 16.935 & 17.256 & 17.763 & 18.119 & 18.489 & 18.117 & 18.220 & 18.418 & 18.529 &39\,000 & sdOB & \\
15:25:34& +09:58:51& 16.207 & 16.471 & 17.065 & 17.353 & 17.829 & 18.102 & 18.331 & 17.977 & 17.887 & 17.634 & 17.625 &31\,000 & sdOB &  \\
15:26:08& +00:16:41& 15.026 & 15.603 & 16.185 & 16.573 & 17.066 & 17.436 & 17.787 & 17.489 & 17.514 & 17.633 & 17.656 &50\,000 &  He-sdOB & (3) J1526+0016\\
15:27:04& +08:02:37& 16.920 & 17.223 & 17.748 & 17.839 & 18.194 & 18.397 & 18.679 & 18.161 & 18.195 & 18.118 & 17.883 &19\,000 &  & \\
15:28:52& +09:31:44& 15.059 & 15.374 & 15.920 & 16.161 & 16.664 & 17.030 & 17.340 & 17.046 & 17.148 & 17.223 & 17.345 &34\,000 &  sdOB &\\
15:35:25& +06:56:52& 15.479 & 15.656 & 16.044 & 16.173 & 16.643 & 16.949 & 17.266 & 16.921 & 17.012 & 17.020 & 17.070 &27\,000 &  & \\
15:38:34& +03:08:13& 15.718 & 15.932 & 16.449 & 16.686 & 17.169 & 17.507 & 17.824 & 17.488 & 17.572 & 17.645 & 17.796 &34\,000 &  & \\
\hline
\end{tabular}
\end{table*}

\subsubsection{Spectroscopic follow-up}

While all hot sd candidates from test region A were already classified, this is not the case for test region B,
where only a few objects are identified in the literature, as explained below. For this reason, we have performed
spectroscopic follow-up for candidates from test region B, which is described in what follows.

\begin{figure*}
   \centering
   \includegraphics[]{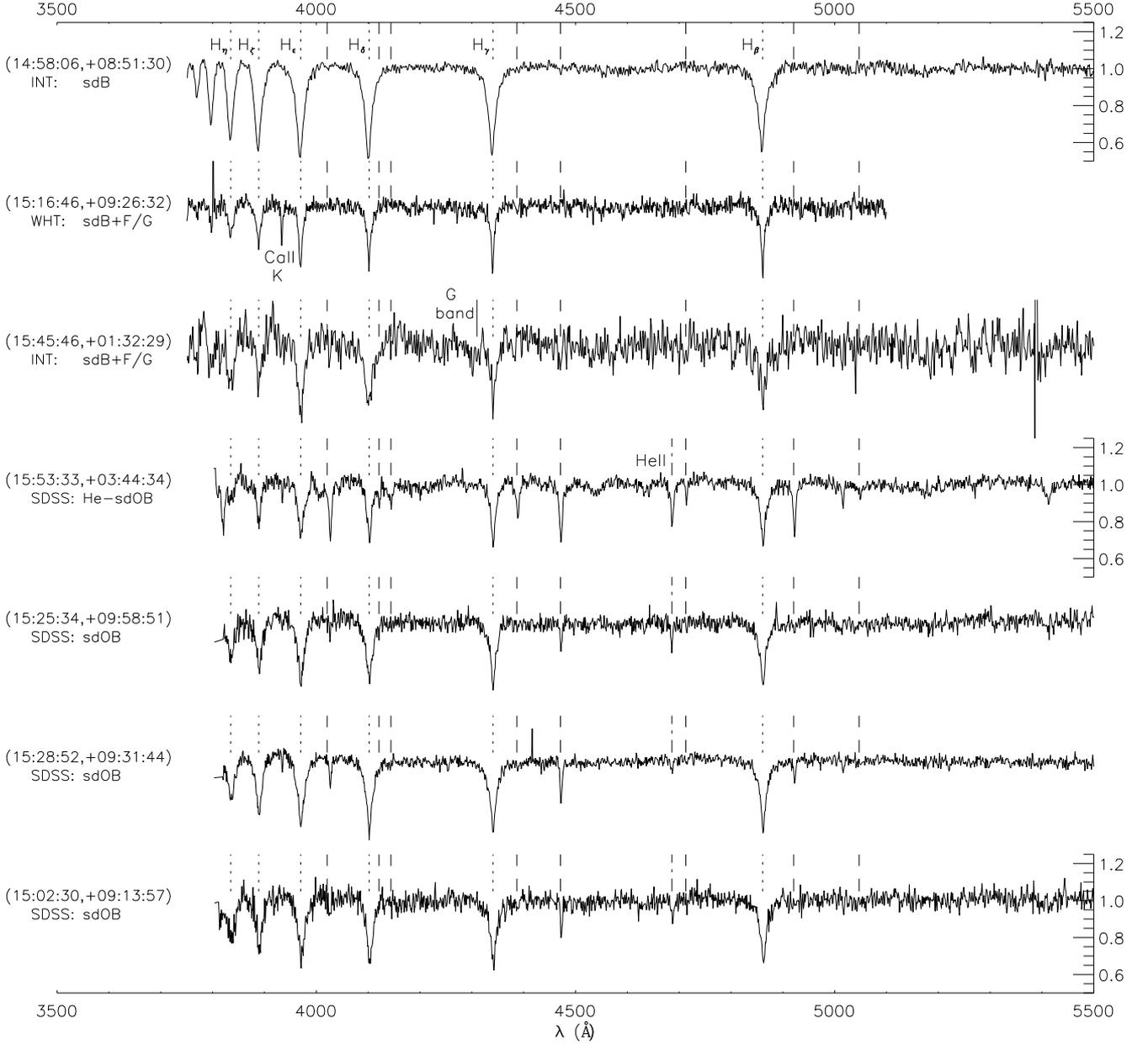}
      \caption{Follow-up spectroscopy gathered at the INT and WHT for some candidates of Table~\ref{tab:candsB2mass},
and SDSS spectra that were available
for some others from Tables~\ref{tab:candsB2mass},\ref{tab:candsBUKIDSS} (last four panels). The first Balmer lines are
indicated as dotted vertical lines; dashed vertical
lines indicate HeI spectral lines; HeII4686 is marked with a dashed-dotted line; the G-band is also indicated when
present.}
         \label{fig:followup}
   \end{figure*}

{\textbf{\it Test region B: {\sc 2mass}-{\sc galex}:}} Five objects from Table~\ref{tab:candsB2mass} have an available
SDSS spectrum. Four of them
(J15104+0409, EGGR\,491, J15513+0649, J15564+01131) are classified as hot subdwarfs in
\cite{ostensennot}, while a visual inspection of the fifth object ((15:53:33,+03:44:34), see Fig.~\ref{fig:followup})
confirms it is also a hot subdwarf. On the other hand, \citet{koester09} classify WD\,1532+033 as a DA
object.

Follow-up spectroscopy of 3 additional objects from Table~\ref{tab:candsB2mass} could be gathered
at the Isaac Newton Telescope (INT) and the William Herschell Telescope (WHT) in La Palma (Spain)
as a filling program.
The IDS spectrograph mounted on the INT was used with the R400B grating providing a resolution of
R$\approx$1400 and an effective wavelength coverage $\lambda\backsimeq$3100--6700\,\AA. The ISIS
spectrograph was used in the case of the WHT with grating R300B on the blue arm
(R$\approx$1600, $\lambda\backsimeq$3100--5300\,\AA ).
Standard {IRAF} packages were used for the data reduction, which included bias subtraction, flatfield correction and
wavelength calibration. The extracted spectra were normalized to
obtain
a raw spectral classification in order to check the success rate of our methodology. The normalized spectra of these
objects are included in Fig.~\ref{fig:followup}.

The first object (14:58:06,+08:51:30) displays a
typical sdB spectrum, with only broad Balmer lines up to a low series number.
The Ca\textsc{ii} H and K lines may indicate an sdB star with an F or G type companion, but may also be caused by a
significant amount of dust along the line of sight. The Ca\textsc{ii} H line, when present, deepens the $H_\epsilon$
line, and is often used as an indicator of sdB+F/G composites. The second object in Fig.~\ref{fig:followup},
(15:16:46,+09:26:32), seems to be a case of such an sdB+F/G composite. The spectrum of (15:45:46,+01:32:29) is noisier, but the
G-band together with the deep $H_\epsilon$ line hints towards a composite nature for this star too. When taken
together with the distinctive red excess seen in the SED (Fig. 7), the composite nature of this object is
quite certain. Modelling this object with two Kurucz components provides an excellent fit, as illustrated in the lower
panel of Fig. 7. The optimum fit is achieved for a hot+cool pair with temperatures 30\,000+5\,000K and a radius ratio
$R_{\rm cool}/R_{\rm hot}\simeq 8$.

The SDSS spectrum of object (15:53:33,+03:44:34) is also included in Fig.~\ref{fig:followup}, which is identified as
a He-sdOB based on the presence of Balmer, HeI and HeII lines.

Note that we use the same classification scheme as in~\cite{ostensen10survey}, in which a distinction is made between
the common He-sdOB stars showing HeI and HeII with almost equal depth, and the hotter and more scarce He-sdOs, with
predominantly HeII lines.

{\textbf{\it Test region B: {\sc ukidss}-{\sc galex}}:} We were unable to perform any follow-up spectroscopy for
candidates in
Table~\ref{tab:candsBUKIDSS} due to their faintness.
However, one object is classified as WD DA by~\cite{mccook99} (PG1422+028), and two objects are identified as hot sds
by~\cite{eisenstein06}
(J1439+0102: sdB; J1526+0016: sdO). Moreover,
three other targets have an SDSS spectrum: (15:02:30,+09:13:57), (15:25:34,+09:58:51), (15:28:52,+09:31:44), and we also include
them in Fig.~\ref{fig:followup} at the bottom.
The three have very similar spectra, displaying some HeI lines: 4027,4471\,\AA~(and 4922\,\AA~in the case of
(15:28:52,+09:31:44)) plus the HeII4686 line (only traces for (15:28:52,+09:31:44)) in addition to the Balmer serie. We classify
all of them as sdOB objects, as indicated in Table~\ref{tab:candsBUKIDSS}.


 

\section{Binary hot subdwarfs in the sample}

In the lists of candidates of both test region A and B
(Tables~\ref{tab:candsKepler},\ref{tab:candsB2mass},\ref{tab:candsBUKIDSS}), we have indicated with an asterisk those
targets for which a bad SED fit is obtained. A bad fit is expected to occur for objects hotter than $T_{\rm
eff}>$50\,000\,K (the hot end of the Kurucz grid used), as this seems to happen for the white dwarfs (19:43:44,+50:04:38)
in Table~\ref{tab:candsKepler} and (15:35:10,+03:11:14) in Table~\ref{tab:candsB2mass}. 

\begin{figure}
   \centering
   \includegraphics[scale=0.5]{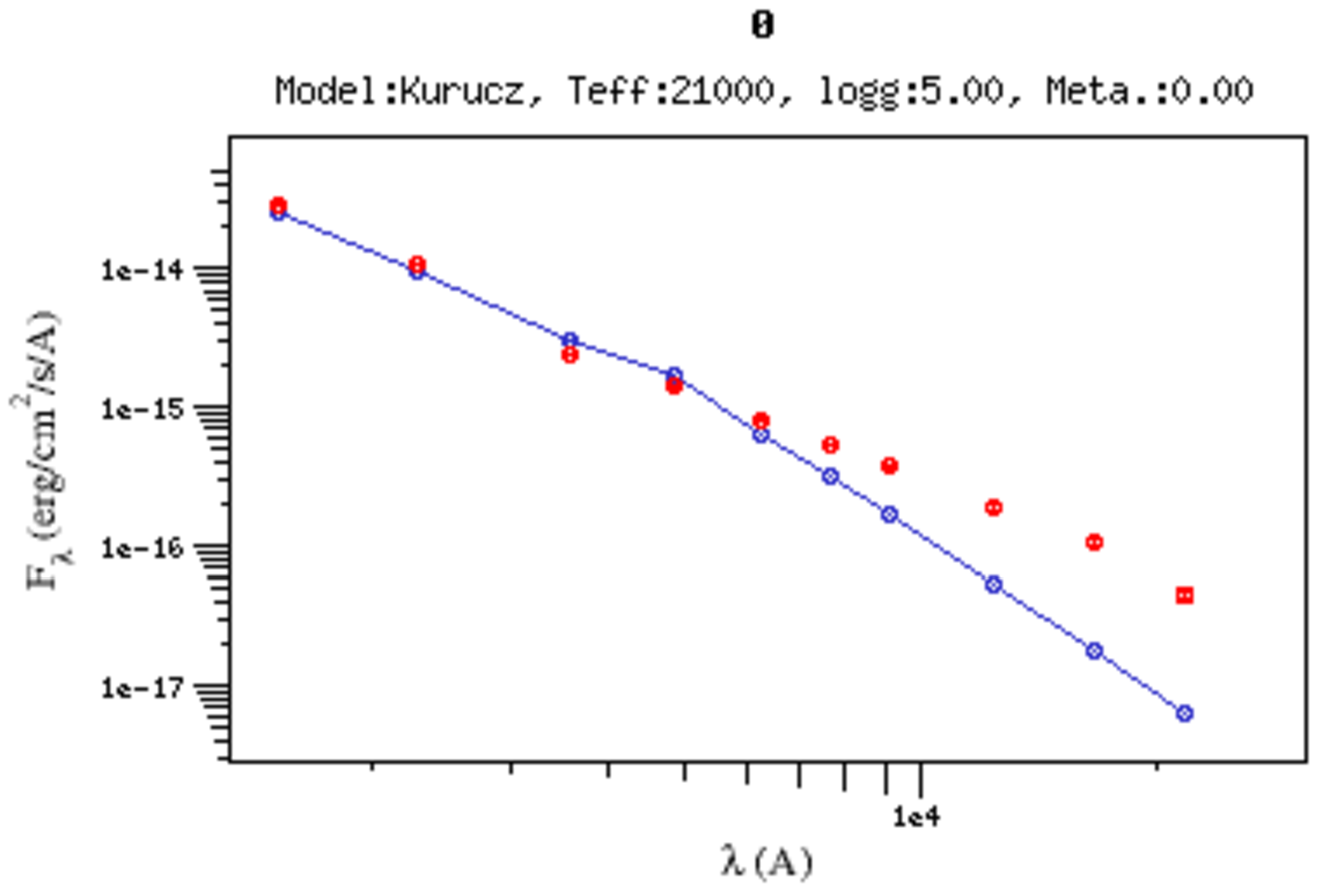}
   \includegraphics[scale=0.8]{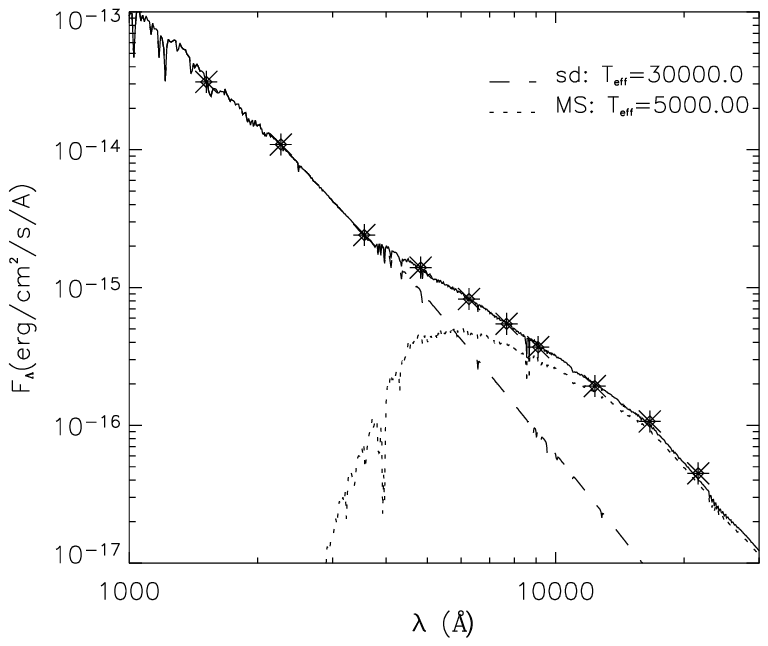}
      \caption{Upper: best VOSA SED fit for the object (15:45:46,+01:32:29). It displays the G-band in the spectrum (see
Fig.~\ref{fig:followup}), and an infrared excess in the spectral distribution. Bottom: best fit obtained considering
two Kurucz components.}
         \label{fig:vosa2}
   \end{figure}

In case a target is in a binary system with a cool companion, its impact at long wavelengths can also cause
a bad SED fit. We have encountered 9 such cases with a more or less clear infrared
excess that may be due to a red companion. For these cases, we have fitted the spectral distribution to a combination of
two Kurucz model atmospheres. In Fig.~\ref{fig:vosa2} we can see how a two-components fit (lower panel) gives a much
better match to
the observed spectral distribution compared to a single-component fit (upper panel) for object (15:45:46,+01:32:29).  The
temperatures of the best hot+cool pair for the nine cases are included in Table~\ref{tab:2compon}. Only
for two targets, the two-components fit did not improve the single-component one performed by VOSA; for them no
temperatures are included in the table and further analyses will be foreseen for these two objects.

\begin{table}
\caption{Two components SED decomposition for objects with a bad fit in
Tables~\ref{tab:candsKepler},\ref{tab:candsB2mass},\ref{tab:candsBUKIDSS} (marked with an asterisk under ''Comments``),
 which may be due to a cool companion. Only for objects (18:42:42 ,+44:04:06) and (19:08:25,+45:08:32) the
two-components fit does not improve the single-component one.}
\label{tab:2compon}
\centering
\begin{tabular}{c|c|c}
\hline\hline
Object & $T_{\rm eff}$ hot & $T_{\rm eff}$ cool \\
\hline
18:42:42  +44:04:06 & --- & --- \\
19:08:25 +45:08:32 & --- & --- \\
19:10:00 +46:40:24 & 37\,000 & 6\,250 \\
19:16:12 +47:49:16 & 22\,000 & 5\,250\\
19:26:51 +49:08:48 & 28\,000 & 5\,000\\
19:40:32 +48:27:23 & 31\,000 & 5\,750\\
15:45:46 +01:32:29 & 30\,000 & 5\,000\\
15:51:20 +06:49:04 & 29\,000 & 5\,500\\
14:15:17 +09:49:26 & 29\,000 & 5\,500\\
\hline                                   
\end{tabular}
\end{table}

\section{Conclusions}

We have developed a methodology to find new uncatalogued hot subdwarfs
making use of large databases such as {\sc galex},
{\sc 2mass}, {\sc ukidss} and SuperCOSMOS. VO tools helped to handle the different queries
and the large output list of candidates. We tested
the methodology using bona-fide input objects from the literature,
which are spectroscopically classified in
the catalogues listed in Section~3.

The {\sc 2mass}-{\sc galex} photometry combination is first used to separate blue
objects from redder ones. A selection criterion given by Eqs.~\ref{eq:fuvk},\ref{eq:fuvnuv} 
effectively retrieves hot sd candidates, along with other UV-excess 
objects: mainly white dwarfs, cataclysmic variables and main
sequence OB stars. 87\% of the input hot sds, while only 10\%, 13\% and 33\% of the input WDs, CVs
and OB stars survive this first filter. Using
reduced proper motions, the hot sd selection further improves,
with very few WDs fulfilling the adopted
proper motion criterion (see Table~\ref{tab:rhee_crit}).

Moreover, the Virtual Observatory Sed Analyzer (VOSA) was used to obtain a rough guess of temperatures for
the candidates. These are computed fitting Kurucz model atmospheres to the
spectral energy distribution, constructed using
all the available photometry for every target. We adopt as hot sds threshold objects with $T_{\rm eff}>$19\,000\,K.
This criterion is specially useful to filter out CVs and OB stars, since, in general, lower
temperatures are obtained by VOSA for these objects (see Fig.~\ref{fig:hist2}).
Targets with a bad fit are retained as good
candidates independently of their temperature estimate, given that a
poor fit may be obtained for stars showing
infrared excess indicative of a cool companion, or for sdOs of high
temperature ($T_{\rm eff}>$50\,000\,K).

After all these filters are imposed, 72\% of the initial hot sds remain, while, on the other hand, only a low fraction
of other spectral types meet the criteria adopted: 3\%, 4\% and 6\% of the initial WDs, CVs and OBs respectively.

We have applied this strategy to two test regions: a $\sim$420\,deg$^2$ region centered on the {\it Kepler}
satellite field of view (test region A), and other at RA:210--240,
DEC:0--10 (test region B). In the {\it Kepler} FoV, 73
objects fulfill the colour and proper motion criteria. However, only 21
have additional KIC $g'r'i'z'$ photometry besides {\sc 2mass} and
{\sc galex}, which is necessary to perform an acceptable SED fit. Thirteen
of them have temperature estimates above
19\,000\,K and two are retained because of its bad fit (see Table~\ref{tab:candsKepler}).  Thanks to the ground-based
support to the {\it Kepler}
mission, spectroscopic follow-up of all
the candidates was available. Thirteen candidates are confirmed
to be hot sds, with only two
targets contaminating the sample: a main sequence B
star, and a white dwarf.

For test region B, we applied the same methodology, and obtained a list of 11 candidates
(see~Table~\ref{tab:candsB2mass}),
3 of them retained due to a bad SED fit. Four of these objects have been recently classified as hot sds by
\cite{ostensennot}, a fifth target is classified as DA by \cite{koester09}, and a sixth candidate (15:53:33,+03:44:34) has
an spectrum in
the SDSS database, which identifies the object as a hot sd. Follow-up spectroscopy could be acquired for 3 other objects
at the INT and WHT (La Palma). The classification is included in Table~\ref{tab:candsB2mass}. 

Furthermore, we repeated the same procedure for test region B, but using {\sc ukidss} instead of the {\sc 2mass}
database.
12 candidates are proposed, out of which 2 are classified as hot sds by~\cite{eisenstein06}, one is labeled as DA by
\cite{mccook99} and 3 are identified as hot sds in this paper based on available Sloan spectra (see
Fig.~\ref{fig:followup}).

In total, we proposed 38 candidates, of which 30 could be spectroscopically classified, and 26 of them
were confirmed to be hot sds. The success rate is thus 87\%. This high percentage confirms the
suitability of our methodology to discover new hot subdwarfs. 

Former surveys described in \S 2 aimed at finding faint blue stars in general and not only hot subdwarfs in
 particular. Thus, it is intrinsically improper to compare our finding rate with the percentage of hot sds found in these
works.
However, they rate a maximum of 53\% of hot sds and demonstrate the difficulty of this task due to the photometric
(and spectroscopic) similitudes among blue objects.

 \cite{ostensen10survey,ostensen11} compile the variety of methods used by several
teams to obtain
uncatalogued blue
compact targets within the {\it Kepler} FoV. Most of the methods are based only on photometric colours and do not
particularly
intend to discern white dwarfs from hot sds, since both classes are of interest from a seismological
point of
view. Their success rate ranges from poor, when only {\sc 2mass} colours are considered, to actually very good, when
using
SDSS filters. The complete sample listed in ~\cite{ostensen10survey,ostensen11}
is formed by 68 sds, 17 WDs, 14 Bs, 2 PNN, 3
CVs and 6 other main sequence stars. Comparing with the number of hot sds candidates proposed in this work,  
only 19 of the 68 hot sds in~\cite{ostensen10survey,ostensen11} have all the necessary data required by our
methodology
({\sc 2mass}, {\sc galex} and proper motion), 17 of which fulfill all our selection criteria. We have retrieved 13 of them;
for the other 4 hot sds, their {\sc galex}
photometry is obtained from the Guest Investigator survey,
which was not public at the time our search was performed, and thus these targets are not listed as candidates in Table~\ref{tab:candsKepler}. 
This serves as a cross-check that: i) a low fraction of hot sds are
rejected by our search method, as 
expected from the tests made in Section 3, and ii) that we will be able to increase our detections as the sky is better covered by the large surveys 
we make use of.

We note that the use of proper motion data was particularly suitable for finding WDs, as described
in~\cite{ostensen10survey,ostensen11}. 
Proper motion information has also been useful to spot
WDs and hot sds by \cite{jimenez10}, whose objective is the
identification of blue high proper motion objects.Combined photometric indexes and proper motion data is also employed by \cite{vennes10} with the intention of finding
new white dwarfs. Their bright sample of candidates contains one single (and already known) WD, plus 15 hot sds (6
already catalogued), 29 main sequence B stars and other 5 blue objects of different nature.

Encouraged by the results in the two pilot regions, and taking advantage of the
Virtual Observatory capabilities, we have initiated a systematic 
search for hot subdwarf stars in the Milky Way, the results of which will be published in a forthcoming paper.

\begin{table*}
\caption{List of the 73 candidates in and around the {\it Kepler} field of view that fulfill the photometric and proper
motion
criteria. FUV, NUV have been taken from the {\sc galex} archive and J,H,K from {\sc 2mass}. 
The KIC number for 21 objects with additional photometry is included. (*) indicates objects meeting all the selection
criteria, detailed in
Table~\ref{tab:candsKepler}.}
\label{tab:candsKepler73}
\centering
\renewcommand{\footnoterule}{}  
\begin{tabular}{c@{ }@{ }c@{ }@{ }c@{ }@{ }c@{ }@{ }c@{ }@{ }c@{ }@{ }c@{ }@{ }c|c@{ }@{ }c@{ }@{ }c@{ }@{ }c@{ }@{ }c@{
}c@{ }@{ }c@{ }@{ }c@{ }@{ }c@{ }@{ }c@{ }}
\hline \hline
RA      & DEC   & FUV     & NUV     &  J & H & K &  KIC & & RA      & DEC   & FUV     & NUV     &  J & H & K & 
KIC \\
(J2000) & (J2000)&         &         &    &   &   & number  & & (J2000) & (J2000)&         &         &    &   &  
&number \\
\hline
18:19:48& +33:22:09& 15.704 & 15.745 & 17.02 & 17.04  & 16.48 & 		&  &  18:50:46&  +51:07:38& 14.416 & 14.134 & 14.01 & 14.10  & 14.17 &   \\
18:20:55& +33:18:47& 16.137 & 16.291 & 17.03 & 16.63  & 17.47 & 		&  &  18:57:58&  +44:40:57& 16.300 & 16.454 & 15.79 & 15.63  & 15.40 &  8544347 \\
18:20:58& +37:07:07& 14.441 & 14.658 & 16.49 & 15.89  & 16.53 & 		&  &  19:03:44&  +47:24:39& 14.991 & 15.303 & 15.45 & 14.94  & 14.93 &   \\
18:21:10& +34:46:45& 13.062 & 14.152 & 15.29 & 15.50  & 15.47 & 		&  &  19:04:35&  +48:10:22& 15.726 & 16.057 & 16.63 & 16.34  & 17.11 &  10784623 (*) \\
18:21:50& +41:51:56& 16.438 & 16.870 & 16.79 & 16.80  & 15.87 & 		&  &  19:05:06&  +43:18:31& 14.208 & 14.391 & 15.81 & 16.06  & 16.23 &  7668647 (*) \\
18:22:43& +43:20:37& 12.865 & 13.110 & 14.40 & 14.52  & 14.50 & 		&  &  19:05:20&  +44:57:59& 17.545 & 17.594 & 16.41 & 15.93  & 17.28 &  8741434 \\
18:23:10& +33:33:55& 14.461 & 14.451 & 15.66 & 15.56  & 15.99 & 		&  &  19:08:25&  +45:08:32& 15.267 & 15.518 & 16.17 & 15.78  & 15.36 &  8874184  (*)\\
18:23:57& +41:29:14& 13.682 & 13.653 & 15.16 & 15.29  & 15.64 & 		&  &  19:08:46&  +42:38:32& 13.983 & 14.490 & 15.90 & 16.05  & 16.02 &  7104168 (*) \\
18:24:00& +51:55:25& 14.362 & 14.801 & 16.25 & 16.08  & 15.94 & 		&  &  19:09:20&  +45:40:57& 14.295 & 14.622 & 14.72 & 14.87  & 14.88 &   \\
18:24:08& +35:16:19& 14.767 & 15.129 & 15.41 & 15.11  & 14.97 & 		&  &  19:09:34&  +46:59:04& 14.695 & 15.009 & 16.36 & 15.71  & 16.61 &  10001893 (*)\\
18:24:34& +38:00:54& 15.723 & 15.811 & 16.39 & 16.94  & 15.74 & 		&  &  19:10:00&  +46:40:25& 14.119 & 14.496 & 13.93 & 13.80  & 13.73 &  9822180  (*) \\
18:24:44& +35:30:42& 14.218 & 14.603 & 16.83 & 16.64  & 17.33 & 		&  &  19:10:24&  +47:09:45& 11.796 & 12.368 & 11.31 & 11.45  & 11.47 &  10130954 \\
18:24:49& +38:51:38& 14.516 & 14.947 & 16.15 & 16.31  & 15.67 & 		&  &  19:14:28&  +45:39:11& 15.002 & 15.339 & 16.02 & 15.57  & 15.68 &  9211123 (*) \\
18:25:19& +40:33:34& 15.010 & 15.139 & 16.26 & 16.43  & 16.76 & 		&  &  19:15:08&  +47:54:20& 13.617 & 14.043 & 13.30 & 13.33  & 13.36 &  10658302 \\
18:26:22& +32:51:08& 14.236 & 14.637 & 15.54 & 16.18  & 15.25 & 		&  &  19:16:12&  +47:49:16& 15.291 & 15.413 & 14.62 & 14.42  & 14.27 &  10593239  (*)\\
18:26:37& +34:37:26& 14.578 & 14.714 & 16.40 & 16.46  & 17.34 & 		&  &  19:20:03&  +49:15:33& 15.392 & 15.567 & 16.76 & 16.29  & 15.84 &   \\
18:27:45& +37:09:31& 16.518 & 16.700 & 16.48 & 16.42  & 16.06 & 		&  &  19:20:18&  +48:06:21& 15.655 & 15.873 & 16.74 & 17.19  & 16.25 &   \\
18:27:55& +36:22:08& 14.688 & 14.938 & 16.60 & 15.86  & 16.13 & 		&  &  19:20:36&  +49:03:16& 14.275 & 14.500 & 13.50 & 13.49  & 13.54 &  11293898 \\
18:28:50& +34:36:50& 14.217 & 14.627 & 16.55 & 15.70  & 16.14 & 		&  &  19:26:52&  +49:08:49& 14.981 & 14.813 & 14.83 & 14.52  & 14.37 &  11350152 (*) \\
18:32:21& +55:03:00& 15.117 & 15.713 & 15.50 & 14.85  & 14.83 & 		&  &  19:30:49&  +54:21:28& 16.309 & 16.531 & 16.62 & 16.28  & 16.57 &   \\
18:33:44& +43:01:06& 15.482 & 15.580 & 16.49 & 16.40  & 16.71 & 		&  &  19:36:33&  +52:45:19& 16.609 & 17.221 & 16.74 & 15.97  & 15.42 &   \\
18:34:54& +44:49:17& 14.607 & 14.726 & 13.85 & 13.63  & 13.51 & 		&  &  19:40:32&  +48:27:24& 13.544 & 13.969 & 13.68 & 13.56  & 13.56 &  10982905 (*) \\
18:35:17& +43:27:30& 14.230 & 14.219 & 14.26 & 14.22  & 14.14 & 		&  &  19:43:44&  +50:04:39& 13.228 & 13.888 & 15.37 & 15.36  & 15.18 &  11822535 (*) \\
18:36:21& +40:59:38& 14.974 & 15.221 & 13.98 & 13.87  & 13.79 & 		&  &  19:44:43&  +54:49:43& 15.654 & 15.748 & 16.11 & 16.01  & 15.83 &   \\
18:36:34& +53:16:57& 12.861 & 13.589 & 12.74 & 12.79  & 12.84 & 		&  &  19:50:24&  +50:09:00& 15.583 & 15.587 & 13.61 & 13.60  & 13.68 &   \\
18:36:42& +41:30:46& 15.252 & 15.477 & 15.08 & 14.91  & 14.97 & 		&  &  19:53:04&  +49:49:34& 15.581 & 15.923 & 16.32 & 16.23  & 15.62 &   \\
18:39:49& +53:00:04& 14.797 & 15.050 & 16.67 & 16.01  & 15.98 & 		&  &  19:53:42&  +49:59:45& 14.713 & 15.329 & 16.47 & 16.27  & 16.47 &   \\
18:40:21& +41:43:15& 15.376 & 15.386 & 15.24 & 15.24  & 15.11 & 		&  &  19:54:52&  +48:22:29& 15.014 & 14.920 & 13.95 & 13.96  & 13.99 &  10937527 \\
18:42:03& +45:31:59& 14.883 & 14.842 & 14.56 & 14.62  & 14.55 & 		&  &  19:56:48&  +53:12:17& 15.413 & 15.941 & 15.27 & 14.76  & 14.54 &   \\
18:42:42& +44:04:05& 16.317 & 16.442 & 16.75 & 16.34  & 15.78 & 8142623 (*)	&  &  19:57:28&  +53:32:03& 13.989 & 14.650 & 14.47 & 14.19  & 14.02 &   \\
18:43:07& +42:59:18& 14.149 & 14.737 & 16.27 & 16.13  & 16.24 & 7335517 (*)	&  &  20:00:01&  +54:09:03& 15.202 & 15.155 & 12.55 & 12.54  & 12.52 &  \\
18:43:56& +45:37:57& 15.009 & 15.204 & 16.75 & 16.90  & 16.79 & 		&  &  20:01:54&  +49:03:54& 15.119 & 15.456 & 16.44 & 15.74  & 17.05 &   \\
18:47:14& +47:41:47& 13.305 & 13.772 & 15.39 & 15.62  & 15.47 & 10449976 (*)	&  &  20:06:33&  +48:33:29& 12.383 & 13.752 & 9.14 & 9.11  & 9.08 &	\\
18:47:46& +50:41:35& 13.877 & 14.061 & 15.24 & 15.45  & 15.94 & 		&  &  20:07:39&  +54:45:16& 17.424 & 17.667 & 16.81 & 16.29  & 15.84 &   \\
18:49:15& +51:16:05& 13.792 & 13.941 & 15.54 & 15.58  & 15.09 & 		&  &  20:09:34&  +55:05:25& 18.663 & 18.538 & 15.76 & 15.52  & 15.44 &   \\
18:50:05& +50:24:22& 15.206 & 15.315 & 14.60 & 14.52  & 14.39 & 		&  &  20:11:52&  +54:50:11& 14.201 & 13.914 & 10.11 & 10.05  & 10.02 &   \\
18:50:17& +43:58:29& 16.501 & 16.387 & 16.71 & 17.15  & 17.00 & 8077281 (*)	&  & &  & &  & &  & &\\
\hline
\end{tabular}
\end{table*}

\begin{acknowledgements}
     This research has made use of the Spanish Virtual Observatory supported from the Spanish MEC 
through grant AyA2008-02156.
R. O., C. R.-L. and A. U. acknowledge financial
support from the Spanish Ministry of Science
and Innovation (MICINN) through grant AYA 2009-14648-02 and from the
Xunta de Galicia
through grants INCITE09 E1R312096ES and INCITE09 312191PR, all of them partially
supported by E.U. FEDER funds. CRL acknowledges an {\em \'Angeles Alvari\~no} contract from the {\em Xunta de
Galicia} and financial support provided from the {\em Annie Jump Cannon} grant of the
Department of Physics and Astronomy of the University of Delaware.\\
R.O. thanks J. Guti\'errez-Soto for a careful reading of the manuscript and fruitful discussions.
The research leading to these results has received funding from the European Research Council under the European Community's Seventh Framework
Programme (FP7/2007-2013)/ERC grant agreement N\textordmasculine227224 ({\sc prosperity}), as well as from the Research Council of K.U.Leuven grant agreement FOA/2008/04.\end{acknowledgements}

%
%
%

\bibliographystyle{aa}
\bibliography{astro-ph}

\end{document}